\documentclass[a4paper,12pt]{article}
 \usepackage[latin1]{inputenc}
\usepackage[T1]{fontenc}
\usepackage{amsmath}
\usepackage{stackrel}
\usepackage{amsfonts}
\usepackage{slashed}
\usepackage{amssymb}
\usepackage{color}
\usepackage{graphicx}
\usepackage{color}
\usepackage{dsfont}
\usepackage{upgreek}
\usepackage{soul}
\usepackage[normalem]{ulem}

 \newcommand{\be}{\begin{equation}}
  \newcommand{\bb}{\begin{equation}}
	 \newcommand{\ee}{\end{equation}}
	 \newcommand{\ba}{\begin{eqnarray}}
\newcommand{\phan}{\vphantom{3^{2^2}}}
		 \newcommand{\ea}{\end{eqnarray}}
		 
		   \newcommand{\bea}{\begin{eqnarray}}
			 \newcommand{\eea}{\end{eqnarray}}
\newcommand{\eqb}{\begin{eqnarray}}
\newcommand{\eqf}{\end{eqnarray}}

\begin{document} \title{Fermion zero modes in the  vortex background of a Chern-Simons-Higgs theory with a hidden sector}
\author{ Gustavo Lozano$^a$, Azadeh Mohammadi$^b$, Fidel A. Schaposnik$^c$
~\\
~\\
{$^a\,$\it \small Departamento de F\'\i sica, FCEYN Universidad de Buenos Aires $\&$ IFIBA CONICET,}\\
{\it \small Pabell\'on 1 Ciudad Universitaria, 1428 Buenos Aires, Argentina}
\\
~\vspace{-3 mm}
\\
{$^b\,$\it \small Departamento de F\'\i sica, Universidade Federal da Para\'\i ba}\\
{\it \small 58.059-970, Caixa Postal 5.008, Jo\~{a}o Pessoa, PB, Brazil}
\\
~\vspace{-3 mm}
\\
{$^c\,$\it \small Departamento de F\'\i sica, Universidad Nacional de La Plata/IFLP/CICBA}
\\{\it \small CC 67, 1900 La Plata, Argentina}
}

\date{}

\maketitle
\begin{abstract}
In this paper we study a $2+1$ dimensional system in which fermions are coupled to the self-dual topological vortex in $U(1)\times U(1)$ Chern-Simons theory, where both $U(1)$ gauge symmetries are spontaneously broken. We consider two Abelian Higgs scalars with visible and hidden sectors coupled to a fermionic field through three interaction Lagrangians, where one of them violates the fermion number.  Using a fine tuning procedure, we could obtain the number of the fermionic zero modes which is equal to the absolute value of the sum of the vortex numbers in the visible and hidden sectors.
\end{abstract}
\section{Introduction}
Zero-eigenvalue modes in the Dirac equation for fermions coupled to a topologically non-trivial background can have remarkable physical consequences in systems belonging to a large domain in physics, going from high energy to condensed matter physics. A typical example in the former case is the so-called QCD  $U(1)_A$  problem which can be explained taking into account the zero modes of Dirac fermions in an instanton background \cite{Coleman}. Also in many condensed matter systems, as for example in superconducting graphene or in topological insulators, zero-mode fermions in a vortex background could play a central role (see \cite{W} and references therein).

Zero modes in $2+1$-dimensional Dirac equations for fermions minimally coupled to an Abelian Higgs model vortex \cite{Abri}-\cite{NO}
 were first studied by Nohl \cite{Nohl} and de Vega \cite{dV}.  Jackiw and Rossi \cite{JR} reconsidered this problem adding a fermion number violating interaction  between fermions and the Higgs field. They  set the fermion mass term to zero but nevertheless a mass was generated dynamically via the fermion-Higgs interaction.
 They found
that the resulting Dirac equation has $|n|$ zero-mode solutions in the $n$-vortex background field.

When a Chern-Simons term is included in the $2+1$-dimensonal Lagrangian, vortex solutions are electrically charged \cite{PaulKhare}-\cite{CuLo}. In this case the search for Dirac equation zero modes  is considerably more involved due to the existence of an electrostatic potential.
Grignani and Nardelli \cite{GN} studied in detail the case in which the vortex background corresponds to a Chern-Simons-Higgs action \cite{H}-\cite{JW} including cases with Yukawa-like interactions between  fermions and the Higgs field. Also, by taking the vortex background as the one arising from a ${\cal N} = 2$ supersymmetric Chem-Simons-Higgs model, Lee et al. found an interesting connection between fermionic and bosonic zero modes \cite{LLM}.

~

Recently,  the study of gauge theories in which a hidden sector is coupled to the Standard Model or its supersymmetric extensions has  received much attention in connection with dark matter, supersymmetry breaking and phenomenological superstring studies (see \cite{1}-\cite{3} and references therein).  In this context self-dual vortex solutions in the bosonic sector of ${\cal N} =2$, $U(1)\times U(1)$ planar gauge theories coupled to Higgs scalars and including Chern-Simons terms  have been considered in \cite{AIST}. There have been also  results concerning the case of a Maxwell-Chern-Simons-Higgs model with  vortex solutions which might have interesting implications in condensed matter problems \cite{Shaposh}.

It is the purpose of the present paper to study the zero-mode Dirac equation for a fermion in the background of the self-dual vortex solutions referred above, namely  a $U(1)\times U(1)$ Chern-Simons  gauge theory, each sector coupled charged scalars with a mixing between both sectors given in \cite{AIST}. We first consider  the case in which  both $U(1)$ gauge symmetries are spontaneously broken so that two distinct topological Chern-Pontryagin numbers characterize the vortex magnetic fluxes and electric charges. We also comment on the case in which one of the Higgs scalars is absent (no symmetry breaking in the corresponding sector) so that there is just one topological number but still magnetic flux and electric charge confined in a flux tube.

\section{The gauge field/Higgs background}
The $2+1$ dimensional bosonic sector of the ${\cal N}=2$ supersymmetric   $U(1) \times U_h(1)$ Chern-Simons gauge theory coupled
to Higgs scalars analyzed in \cite{AIST} reads
\be
L=\kappa\epsilon^{\mu\nu\rho}A_\mu \partial_\nu A_\rho  +  \kappa_h\epsilon^{\mu\nu\rho}C_\mu \partial_\nu C_\rho + 2\xi\epsilon^{\mu\nu\rho}A_\mu \partial_\nu C_\rho + |D_\mu[A]\phi|^2 + |D_\mu[C]\eta|^2 - V_\text{eff}[\phi,\eta]
\label{lagrangian}
\ee
Here
\be
D_\mu^A\phi = \partial_\mu \phi - iq A_\mu\phi \; , \;\;\;\;\;
D_\mu^C\eta = \partial_\mu \eta - iq_h C_\mu\eta
\label{derivadas}
\ee
where $q$  is the charge coupling the $U(1)$   gauge $A_\mu$ to the Higgs scalar $\phi$ while $q_h$
plays the same role for the $U_h(1)$ gauge field $G_\mu$ and the scalar $\eta$.
Concerning   $V_{eff}[\phi,\eta]$, supersymmetry forces the sixth order potential to take the form
\bea
 V_{eff} &=&  {\frac{1}{4(\kappa\kappa_h-\xi^2)}}
 \left(\left( {\kappa_h q} (|\phi|^2- \phi_0^2)+  {\xi q_h} (|\eta|^2- \eta_0^2)\right)^2 e^2|\phi|^2 \right.\\
&&  +\left.
 \left( {\kappa q_h} (|\eta|^2- \eta_0^2)+ {\xi q} (|\phi|^2- \phi_0^2)\right)^2  g^2|\eta|^2
 \right) \ ,
 \eea
where $\xi^2<\kappa \kappa_h$. Note that the Lagrangian includes a gauge mixing term (with parameter $\xi$) which also implies a Higgs portal
mixing the two scalars in the symmetry breaking potential.

The vortex configurations ansatz  takes the following form
 \bea
&& \phi= \phi_0 f(r) e^{in\theta},\,\,\,\,\,\,\, A_\varphi=-\frac{ A(r)}{qr}, \,\,\,\,\,\, A_r=0 \nonumber\\
&& \eta= \eta_0 h(r) e^{ik\theta},\,\,\,\,\,\,\, C_\varphi=-\frac{ C(r)}{q_h r}, \,\,\,\,\,\, C_r=0
\label{ansatz}
\eea
where $n$ and $k$ are the winding numbers associated to each Higgs scalar.

The Bogomolny equations for static configurations associated to Lagrangian \eqref{lagrangian}  can be easily obtained  from vanishing of the SUSY charges acting on physical states  of  the extended supersymmetric model \cite{AIST}. They read
\bea
D_1^A\phi&=&\mathrm {\rm  sgn}(n) iD_2^A \phi \nonumber \\
D_1^C\eta &=&\mathrm {\rm  sgn}(k) iD_2^C \eta \nonumber \\
  B &=&  -\mathrm {\rm  sgn}(n) \frac{q^2|\phi|^2}{2(\kappa \kappa_h-\xi^2)(\kappa - \xi)}\left(\vphantom{\frac{\phi^2}{\xi^2}} {q\kappa_h} (|\phi|^2- \phi_0^2)+ {\xi q_h} (|\eta|^2- \eta_0^2)\right) \ ,\nonumber\\
  B_h &=&  -\mathrm {\rm  sgn}(k) \frac{q_h^2|\eta|^2}{2(\kappa \kappa_h-\xi^2)(\kappa_h - \xi)}\left( \vphantom{\frac{\phi^2}{\xi^2}} {q_h\kappa} (|\eta|^2- \eta_0^2)+ { \xi q} (|\phi|^2- \phi_0^2)\right) \ ,
  \label{22}
\nonumber \\
 A_0 &=&   -\mathrm {\rm  sgn}(n) {\frac{1}{ (\kappa\kappa_h-\xi^2)}}\left( {\kappa_h q} (|\phi|^2- \phi_0^2)+  {\xi q_h} (|\eta|^2- \eta_0^2)\right) \ ,\nonumber\\
 C_0 &=& -\mathrm {\rm  sgn}(k)  {\frac{1}{(\kappa\kappa_h-\xi^2)}}\left( {\kappa q_h} (|\eta|^2- \eta_0^2)+ {\xi q} (|\phi|^2- \phi_0^2)\right)  \ , \label{21}
 \eea
where the magnetic fields are defined as $B=F_{12}[A]$ and $B_h=F_{12}[G]$ and $\mathrm {\rm  sgn}(n)=\frac{|n|}{n}$. The numerical solutions, which also solve the second order field equations are described in \cite{AIST}.

\section{The Dirac zero-mode equation}
Extending the proposal in \cite{GN} for  the ordinary Chern-Simons-Higgs self-dual model, we shall consider the following Dirac Lagrangian for the case in which a hidden sector is included
\be
L_D = \bar \psi\left(\vphantom{3^{2^2}} \gamma^\mu  (i \partial_\mu - e A_\mu - e_h C_\mu) - m
\right)\psi + L_{\phi\eta} +  L_\phi + L\eta \ ,
\label{12}\ee
with
\bea
L_{\phi\eta} &=&  -\frac{i}2 g\phi\eta
\bar\psi  \psi^c + \frac{i}2 g^*\phi^*\eta^* \bar \psi^c   \psi
\label{unoss} \ ,\\
L_\phi &=& - g_\phi |\phi|^2 \bar\psi \psi \label{doss} \ ,\\
L_\eta &=& - g_\eta |\eta|^2 \bar\psi \psi \ . \label{tress}
\eea
We are working in $2+1$ dimensions and hence the lowest dimension  representation for Dirac $\gamma$-matrices is two. They can be chosen as
\be
 \gamma_0 = \sigma^3 \; , \;\;\;\;\; \gamma_1 = i \sigma^ 2\; , \;\;\;\;\; \gamma_2 = -i \sigma^1
 \ee
where $\sigma_i$ are the Pauli matrices. This leads to $ \alpha$-Dirac matrices of the form
 $\vec \alpha = (\sigma^1,\sigma^2)$ and $\beta = \sigma^3$. Concerning the charge conjugated spinor $\psi^c$, it  is related to $\bar \psi$ according
to
\be
\psi^{c\,\alpha} = C^{\alpha\beta} \bar\psi^\beta
\ee
with $C^{\alpha\beta}$ the charge conjugation matrix.

All three interaction Lagrangians, Eqs.\,\eqref{unoss}-\eqref{tress}, correspond to power counting renormalizable intersections in $2+1$ dimensions. It should be also noted that similar fermion/scalar interactions arise in the  supersymmetric extensions of Abelian Higgs models \cite{LLW}. Also note that in the presence of interaction \eqref{unoss}, fermion number which corresponds to an invariance under the global transformation $\psi \to \exp(i\alpha)\psi$ is violated.

In view of the covariant derivatives definitions given in  (\ref{derivadas}),  gauge invariance implies that
  gauge and matter fields should change according to
\be
\begin{array}{cc}
  A_\mu \stackrel{\Lambda}{\longrightarrow} A'_\mu = A_\mu - \partial_\mu \Lambda(x)
   &
  C_\mu \stackrel{\Lambda_h}{\longrightarrow} C'_\mu = C_\mu - \partial_\mu \Lambda_h(x)\\
         \phi \stackrel{\Lambda}{\longrightarrow} \phi' = \exp(iq  \Lambda(x)) \phi & \eta \stackrel{\Lambda_h}{\longrightarrow}  \eta' = \exp(iq_h \Lambda_h(x)) \eta\\
          \psi \stackrel{\Lambda}{\longrightarrow} \psi' = \exp(ie  \Lambda(x)) \psi & \psi \stackrel{\Lambda_h}{\longrightarrow}  \psi' = \exp(ie_h \Lambda_h(x)) \phi
\end{array}
\ee
In the presence of the fermion number violating interaction Lagrangian \eqref{unoss}, the matter-gauge fields couplings should obey the following relations in order to have a $U(1)\times U(1)_h$ gauge invariant theory \cite{JR}
\be
q = 2e \;, \;\;\;\; \;\;\;\; \;\;\;\; q_h = 2e_h
\ee
Interaction Lagrangians $L_\phi$ and $L_\eta$ do not impose any relation between charges.

Notice that in the present case, like in \cite{GN}, the fermion-scalar interaction is quadratic in the scalar field and the coupling constant is dimensionless, which contrasts with \cite{JR} where the coupling is lineal in the scalar. On the other hand, like in \cite{JR}, the fermion number violating term involves the two scalar fields and consequently the scalar charge is twice the electron charge, unlike the case in \cite{GN} where the scalar charge is the same as the fermion charge.

The Dirac equation following Lagrangian \eqref{12} takes the form
\begin{align}
&\left(i\partial_0 - eA_0(\vec x) - e_hC_0(\vec x)\right)\psi(\vec x,t) =
\left(
\phan -\vec \alpha.\left(i\vec\nabla + e\vec A + e_h\vec C\right) + \beta m
\right)\psi(\vec x,t)  \nonumber\\& - g f(r) \phi_0 \eta_0 h(r)\exp(i(n+k)\varphi)\sigma^2 \psi^*(t,\vec x)
+\left(\phan g_\phi \phi_0^2f^2(r) + g_\eta \eta_0^2h^2(r)\right) \sigma^3\psi(t,\vec x) = 0 \ .
\end{align}

Following \cite{JR}-\cite{GN} we make the two phases ansatz  in order to factor out time dependence
\be
\psi(t,r,\varphi) = \exp(-iEt) \Psi^+(r,\varphi) + \exp(iEt) \Psi^-(r,\varphi) \ .
\ee
Now, for the zero-mode solutions the two equations collapse to one which takes the form
\begin{align}
&\left(
\phan -\vec \alpha.\left(i\vec\nabla + e\vec A + e_h\vec C\right) + eA_0 + e_h C_0 +\beta m
\right)\Psi(r,\varphi )  \nonumber\\& - g f(r)\phi_0 \eta_0 h(r)\exp(i(n+k)\varphi)\sigma^2 \Psi^*(r,\varphi)
+\left(\phan g_\phi \phi_0^2f^2(r) + g_\eta \eta_0^2h^2(r)\right) \sigma^3 \Psi(r,\varphi) = 0
\label{isi}
\end{align}
In terms of upper and lower components of the spinor field $\Psi$, $\Psi = \left(
\begin{array}{l}
\Psi_U\\
\Psi_L
\end{array}
\right)$,
the above equation can be written in the form of the coupled system,
\begin{align}
& (eA_0 + e_hC_0 + m) \Psi_U - i e^{-i\varphi}\left(\partial_r - \frac{i}r\partial_\varphi +\frac12( A(r) +  C(r))
\right) \Psi_L\nonumber\\
& + ig f(r) \phi_0 \eta_0 h(r)\exp(i(n+k)\varphi)\Psi_L^*(r,\varphi)
+\left(\phan g_\phi \phi_0^2f^2(r) + g_\eta \eta_0^2h^2(r)\right) \Psi_U(r,\varphi)  =0 \ ,
\end{align}
and
\begin{align}
& (eA_0 + e_hC_0 - m) \Psi_L - i e^{i\varphi}\left(\partial_r + \frac{i}r\partial_\varphi - \frac12(A(r)+C(r))
\right) \Psi_U\nonumber\\
& - ig f(r) \phi_0 \eta_0 h(r)\exp(i(n+k)\varphi) \Psi_U^*(r,\varphi)
-\left(\phan g_\phi \phi_0^2f^2(r) + g_\eta \eta_0^2h^2(r)\right) \Psi_L(r,\varphi)  =0 \ .
\end{align}
Proposing the two-phase ansatz as
\begin{align}
\Psi_U&=U_U e^{im\varphi}+V_U e^{i(n+k-m-1)\varphi} \ , \nonumber \\
\Psi_L&=U_L e^{i(m+1)\varphi}+V_L e^{i(n+k-m)\varphi} \ ,
\end{align}
with $m \in \mathbf{Z} $, we obtain the following set of coupled equations
\begin{align}
&\left(eA_0 + e_hC_0 + m\right) U_U - i \left[\partial_r+\frac{m+1}{r}+ \frac12\left(A(r) +C(r)\right)\right] U_L\nonumber \\
&+igf(r)\phi_0 \eta_0 h(r) V^*_L+\left(g_\phi \phi_0^2 f^2(r)+g_h \eta_0^2 h^2(r)\right) U_U=0 \ , \label{19a}
\\
&\left(eA_0 + e_hC_0 + m\right) V_U - i \left[\partial_r+\frac{n+k-m}{r} + \frac12\left(A(r) +C(r)\right)\right] V_L\nonumber \\
&+igf(r)\phi_0 \eta_0 h(r) U^*_L+\left(g_\phi \phi_0^2 f^2(r)+g_h \eta_0^2 h^2(r)\right) V_U=0 \ ,
\label{20a}
\\
&\left(eA_0 + e_hC_0 - m\right) U_L- i \left [\partial_r-\frac{m}{r}-\frac12\left(A(r) +C(r)\right)\right ] U_U\nonumber \\
&-igf(r)\phi_0 \eta_0 h(r) V^*_U-\left(g_\phi \phi_0^2 f^2(r)+g_h \eta_0^2 h^2(r)\right) U_L=0
\label{21a} \ ,
\\
&\left(eA_0 + e_h C_0 - m\right) V_L - i \left[\partial_r-\frac{n+k-m-1}{r}-\frac12\left(A(r) +C(r)\right) \right] V_U\nonumber \\
&-igf(r)\phi_0 \eta_0 h(r) U^*_U-\left(g_\phi \phi_0^2 f^2(r)+g_h \eta_0^2 h^2(r)\right) V_L=0 \ .
\label{22a}
\end{align}

Defining the following relations
 \be
 {\cal M} \equiv \frac{ q^2 \phi_0^2}{|\kappa|} \; , \;\;\;\;
 {\cal N} \equiv \frac{  q_h^2 \eta_0^2} {|\kappa_h|}
 \ee
one can rewrite the zeroth components of the gauge fields in Eq.\eqref{21} in the form
 \bea
 eA_0 &=&   {\rm  sgn}(n) \left(\frac{|\kappa|\kappa_h e}{q(\kappa\kappa_h-\xi^2)}{\cal M} (1 - f^2)+
 \frac{|\kappa_h|\xi e }{q_h(\kappa\kappa_h-\xi^2)}{\cal N} (1 - h^2)\right) \ , \nonumber\\
 e_hC_0 &=&  {\rm  sgn}(k) \left( \frac{|\kappa_h|\kappa e_h}{q_h(\kappa\kappa_h-\xi^2)} {\cal N} (1 - h^2)+ \   \frac{|\kappa|\xi e_h}{q(\kappa\kappa_h-\xi^2)}  {\cal M} (1- f^2)  \right) \ . \label{21NM}
 \eea
 As a result,  we obtain
\begin{align}
eA_0+e_hC_0 =\left(|\kappa|\frac{\mathrm {\rm  sgn}(n) \kappa_h e +\mathrm {\rm  sgn}(k) \xi e_h}{(\kappa\kappa_h-\xi^2)}\frac{\cal M}{q} (1 -f^2)
+          |\kappa_h|   \frac{\mathrm {\rm  sgn}(k) \kappa e_h +\mathrm {\rm  sgn}(n) \xi e }{ (\kappa\kappa_h-\xi^2)}\frac{\cal N}{q_h}(1 - h^2)\right)
\end{align}
Now, calling
\bea
l &\equiv& \frac{\kappa\kappa_h-\xi^2}{\kappa\left[\mathrm {\rm  sgn}(n)\kappa_h e +\mathrm {\rm  sgn}(k)\xi e_h\right]} q \ , \nonumber\\
l_h &\equiv& \frac{\kappa\kappa_h-\xi^2}{\kappa_h\left[\mathrm {\rm  sgn}(k)\kappa e_h +\mathrm {\rm  sgn}(n)\xi e\right]} q_h \ ,
\eea
we finally get
\be
eA_0+e_hC_0 = \left(\mathrm {\rm  sgn}(\kappa) \frac{\cal M}{l} (1 -f^2) +
              \mathrm {\rm  sgn}(\kappa_h) \frac{\cal N}{ l_h} (1 -h^2)\right) \ ,
\ee
which can be written as the extension of Grignani-Nardeli result to the $U(1)\times U(1)$ model if one defines
\be
e\hat A_0 \equiv \mathrm {\rm  sgn} (\kappa) \frac{\cal M}{l}  \;, \;\;\;\;
e_h\hat C_0 \equiv  \mathrm {\rm  sgn}(\kappa_h) \frac{\cal N}{l_h}
\ee
It results in
\be
eA_0+e_hC_0 = e\hat A_0 (1 -f^2) + e_h \hat C_0 (1 -h^2) \ .
\ee

Let us define $I_0$ as
\bea
I_0  \equiv  \left(\vphantom{\frac12}(e\hat A_0 -  g_\phi \phi_0^2) (1 - f^2) +
                              (e_h\hat C_0 -  g_\eta \eta_0^2) (1 - h^2)
 +M_f\right) \ ,
 \eea
 with
 \be
M_F \equiv  m + g_\phi \phi_0^2 +  g_\eta \eta_0^2 \ .
\label{conditionm}
 \ee
Then,   the terms containing   an $U_U$  factor  in Eq.\,\eqref{19a} and a $V_U$ factor in Eq.\,\eqref{20a} take the form  $I_0 U_U$ and $I_0 V_V$, respectively.
Analogously, the terms containing $U_L$ and $V_L$ in Eqs.\,\eqref{21a}-\eqref{22a} are of the form
$J_0 U_l$ and $J_0 V_L$ with
\be
J_0
\equiv   \left(\vphantom{\frac12}(e\hat A_0 +  g_\phi \phi_0^2) (1 - f^2) +
                              (e_h\hat C_0 +  g_h \eta_0^2) (1 - h^2)
 -M_f\right) \ .
 \ee
 Then, in order to find non-trivial zero-mode solutions the analysis in \cite{GN} for the case of just one sector can be extended to the present model leading to the following fine tuning conditions
 \bea
  M_f &=& 0 \ , \\
  \hat A_0 &=& \pm\frac{g_\phi}{e} \phi_0^2  \label{34} \ ,\\
 \hat C_0 &=& \pm\frac{g_h}{e_h} \eta_0^2 \ , \label{344}
 \label{condi}
 \eea
 so that either $I_0 = 0$ or $J_0 = 0$, depending on the sign choice.

Let us first consider the case in which the signs   in Eq.\,\eqref{34} and in Eq.\,\eqref{344} are negative.  Under these conditions, the upper components, $U_U$ and $V_U$ completely decouple from the lower components, so that both of them can be taken equal to zero. Then
 Eqs.\eqref{19a}-\eqref{20a} take the simple form
 \bea
     \left(\partial_r+\frac{m+1}{r}+ \frac12\left( A(r) +C(r)\right)\right) U_L  &=&
 gf(r)\phi_0 \eta_0 h(r) V^*_L  \ , \nonumber\\
 \left(\partial_r+\frac{n+k-m}{r}+ \frac12\left(A(r) +C(r)\right)\right) V_L  &=&
 gf(r)\phi_0 \eta_0 h(r) U^*_L \ .
 \eea
 These equations can be simplified by defining new functions $\tilde U_L$ and $\tilde V_L$ through the relations
 \bea
 U_L &=& \exp\left( -\frac12 \int_0^r dr' (A(r') +C(r')) \right) r^{-m-1} \tilde U_L \ ,
 \nonumber\\
 V_L &=& \exp\left( -\frac12 \int_0^r dr' (A(r') +C(r')) \right) r^{m - (n+k)} \tilde V_L \ ,
 \label{38}
 \eea
 so that one now has
 \bea
 \partial_r \tilde U_L &=& g\phi_0\eta_0 f(r)h(r) \tilde V_L^* r^{(2m  -(k+n) +1)} \ ,
 \nonumber\\
 \partial_r \tilde V_L &=& g\phi_0\eta_0 f(r)h(r) \tilde U_L^* r^{(-2m  +(k+n) -1)} \ .
 \eea
 Because of the properties of the vortex solutions, the exponentials in Eq.\,\eqref{38} go to 1 at the origin and to $r^{-(n+k)/2}$ at infinity.

One can now obtain  a set of real equations by writing
\be
\tilde U_L = \exp(i\omega_L) \hat U_L \; , \;\;\;\; \tilde V_L = \exp(-i\omega_L) \hat V_L
\ee
 with $\omega_L$ an arbitrary phase and $\hat U_L$ and $\hat V_L$ solutions of the equations
 \bea
 \partial_r \hat U_L &=& g \phi_0\eta_0 f(r)h(r) \hat V_L r^{(2m - (n+k) +1)} \ , \nonumber\\
 \partial_r \hat V_L &=& g \phi_0\eta_0 f(r)h(r) \hat U_L r^{(-2m + (n+k) - 1)} \ .
 \eea
 From these equations we see that $\hat U_L$ and $\hat V_L$, apart from a power behavior, have an exponential behavior  $\exp(\pm|g\phi_0\eta_0| r)$.  In order to have regular zero-mode solutions, $U_L$ and $V_L$ should be well-behaved at the origin as well as at infinity. It follows that
 \bea
  U_L  &\stackbin[{\rm small~r}]{\sim}{}& {r^{-m - 1}} \; ,  \;\;\; \;\;\; r^{m - (n+k) +  |n| + |k| + 1} \nonumber\\
  V_l &\stackbin[{\rm small~r}]{\sim}{}& r^{m - (n+k)} \; , \;\;\; r^{-m + |n|+| k|}
  \label{sim}
 \eea
All the solutions to this equations are regular provided the following inequalities hold
\be
(n + k) \leq m \leq -1 \ ,
\ee
which in turn implies  $n+k <0$ and therefore the number of zero modes is $|n+k|$.

If one chooses $A_0$ and $C_0$ with   positive signs instead of those chosen in the analysis above, the analysis goes the same. One can take $U_L=V_L = 0$ and instead of
Eq.\,\eqref{sim} we have
 \bea
  U_L  &\stackbin[{\rm small~r}]{\sim}{}& {r^{m}} \; ,  \;\;\; \;\;\; r^{-m + (n+k) +  |n| +|k|} \nonumber\\
  V_l &\stackbin[{\rm small~r}]{\sim}{}& r^{- m + (n+k) - 1} \; , \;\;\; r^{m + |n| + |k| + 1}
  \label{sim2}
 \eea
 leading to the inequalities
 \be
n+k-1 \geq m \geq0 \ ,
\ee
which can only be satisfied for positive $n+k$. Therefore, it results in the existence of $(n+k)$ zero modes.

The final form of the zero-energy eigenfunctions for the case $n+k< 0$ is
\begin{align}
\psi_{(n+k)<0}(\vec r)  &=  \exp\left(-\frac12 \int_0^r dr' \left(A(r') +C(r')\right)  \right)
\nonumber\\
 & \times\left( e^{i((m+1)\varphi + \omega_L)}r^{-m-1}  \hat{U}_L  + e^{i( (n+k-m)\varphi - \omega_L)} r^{m - (k+n)}\hat{V}_L \right) \left(
 \begin{array}{l}
 0\\
 1
 \end{array}
 \right) \ ,
\end{align}
while for $n+k >0$ one has
\begin{align}
\psi_{(n+k)>0}(\vec r)  &=  \exp\left(-\frac12 \int_0^r dr' \left(A(r') +C(r')\right)  \right)
\nonumber\\
& \times\left( e^{i(m\varphi + \omega_U)}r^{m}  \hat{U}_U  + e^{i( (n+k-m - 1)\varphi - \omega_U)} r^{(k+n) - m - 1}\hat{V}_U \right) \left(
 \begin{array}{l}
 1\\
 0
 \end{array}
 \right) \ .
 \label{ultima}
\end{align}

One can see that as a result of fermion number violation, $\psi$ is an eigenstate of particle conjugation as defined in \cite{JR},
\be
\sigma^3 \psi = {\rm sgn}(n+k) \psi
\ee

One can also find zero modes in the absence of fermion-number violating terms. In this case $q/e$ and $q_h/e_h$ are not fixed by gauge invariance so that, as in refs. \cite{Nohl}-\cite{dV} for just a visible sector, the zero modes will depend on this ratio.

\section*{Summary and discussion}

In this paper we have found all zero modes of fermions in the background of vortex solutions of a Chern-Simons-Higgs model with visible and hidden gauge and Higgs fields. Apart from their quantized magnetic flux, the vortices are electrically charged and they are  solutions of self-dual equations found in  \cite{AIST} by considering the ${\cal N}=2$ supersymmetric extension of the model.

What we have analyzed in this paper is the Dirac equation for a single fermion field that couples to the vortex background fields of the two sectors. It could be of interest to instead consider the zero modes of the ${\cal N}=2$ supersymmetric model with fermions in the two sectors, a $U(1)\times U(1)$ extension of the analysis presented in  \cite{LLM} for just one sector, where a simple connection between all independent fermion and boson zero modes is established through a formula that converts every fermion zero mode into a correspondent bosonic zero mode. Analogous relation should exist when the two $U(1)$ sectors are considered.

 Another direction to explore has to do  with possible condensed matter applications. In this respect, we have seen that zero-modes as given in Eq.\,\eqref{ultima} can be interpreted as their own antiparticles and hence as discussed in \cite{W} they can play a relevant role in describing  certain quantum Hall states and some exotic superconductors.  Also in connection with superconductivity it was shown in \cite{AIST} that when one of the two $U(1)$ symmetries remains unbroken (for example when no Higgs field in the corresponding sector is present)  the magnetic and electric fields are proportional (and so are the magnetic flux and the electric charge). In this case the model makes contact with the one considered by Anber et al  \cite{Shaposh}, except that no Maxwell term is included in any of the two sectors. It could be interesting to see whether the condensed matter applications discussed in \cite{Shaposh}, in particular the possibility of superconductivity at any temperature, can be analyzed from the fermionic zero modes side.

~

\noindent{\bf{Acknowledgments:}}
A.M. thanks Conselho Nacional de
Desenvolvimento Cient\' ifico e Tecnol\' ogico (CNPq) for the financial support within the frame
of Grant No.\,501795/2013-8. F.A.S. is associated to CICBA and financially supported by PIP-CONICET,
PICT-ANPCyT, UNLP and CICBA grants. G.S.L is finacially supported by PIP-CONICET and UBA.

\end{document}